\documentclass[twocolumn,twoside,slac_two]{revtex4}
\usepackage{graphicx}
\usepackage{fancyhdr}
\begin{document}

\title{Three-dimensional Two-Layer Outer Gap Model: the Third Peak of Vela Pulsar}

%

\author{Y. Wang, J Takata $\&$ K.S. Cheng}
\affiliation{Department of Physics, University of Hong Kong, Pokfulam Road, Hong Kong}
%

\begin{abstract}
We extend  the two-dimensional two-layer outer gap model to a three-dimensional geometry and use it to study the {{}high-energy emission} of the Vela pulsar.
We apply this three-dimensional two-layer model to the Vela pulsar and compare the model
light curves, the phase-averaged spectrum and the phase-resolved spectra with the recent $Fermi$ observations, which also reveals the existence of the  third peak between two main peaks.
The phase position of the third peak moves with
the photon energy, which cannot be explained by the geometry of magnetic field structure and
 the caustic effect of the photon propagation.
We suggest that the existence of the
third peak and its energy dependent movement results from the azimuthal structure
of the outer gap.

\end{abstract}

\maketitle

\thispagestyle{fancy}


\section{INTRODUCTION}
The $Fermi$'s observation of the Vela Pulsar \cite{obs_Vela} shows that, in its light curve, as the energy increases, a third peak appears at the trailing part of the first peak and shifts towards the second peak, which makes this known $\gamma$-ray pulsar before $Fermi$'s era more special.

We want to use our two-layer Outer Gap model \cite{2D} to explain the strange third peak and its movements. Our two-dimensional two-layer model has explained the phase averaged spectra of the mature pulsars, including the Vela pulsar, in the first catalogue of $\gamma$-ray pulsars of $Fermi$ \cite{catalogue}. However, to study the pulse profile, a three-dimensional model is needed. Therefore, we extend the two-layer model to a three-dimensional magnetic field. Here we present our simulations of the energy dependent light curves and the explanation of the moving third peak of Vela Pulsar, by three-dimensional two-layer Outer Gap model.

\section{THE THREE-DIMENSIONAL TWO-LAYER OUTER GAP}

\begin{figure}[h]
\centering
\includegraphics[width=65mm]{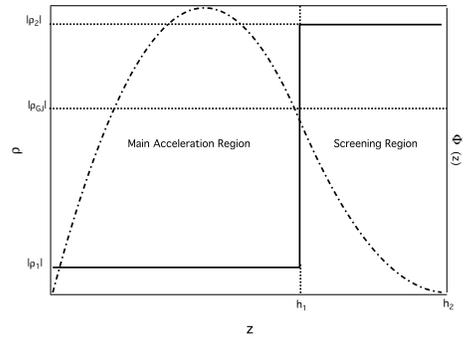}
\caption{The simplified distribution of the charge density (solid line) and the corresponding accelerating potential (dot-dashed line) of the two-layer outer gap.} \label{struct}
\end{figure}

In this study, the three-dimensional rotating vacuum dipole field is adopted.
The $\gamma$-ray emissions come from the region with strong accelerating electric field, which extends
above the last-open filed lines and between the
null charge surface of the Goldreich-Julian charge density \cite{GJ}
and the light cylinder. This region has a two-layer structure. As shown in Figure~\ref{struct}, in the trans-field direction of the magnetic field,
the outer gap
 can be divided into two parts:

\begin{itemize}
\item[1] the main acceleration region at the lower part
of the outer gap, where the charge density is assumed to be $\sim$ 10~\% of the
Goldreich-Julian value and a strong electric field is accelerating
the particles to emit GeV photons via the curvature process,
\item[2] the screening region around the upper boundary, where
 the growth of the main acceleration region in the trans-field direction is stopped
 by the pair-creation processes.
\end{itemize}

And we use a simple step function to approximate the distribution of the charge density
in the trans-field direction in the poloidal plane {(the plane where the field lines have same polar angle $\phi_p$)}as
\begin{equation}
\rho(x, z, \phi_p)=\left\{
   \begin{array}{ccc}
            \rho_1(x, \phi_p), &$if$& 0\leq{}z\leq{}h_1(x, \phi_p)\\
            \rho_2(x, \phi_p), &$if$& h_1(x, \phi_p)<z\leq{}h_2(x, \phi_p)
   \end{array},\right.
\end{equation}
where $x$, $z$ and $\phi_p$ represent coordinates along the magnetic field line,
the height measured from the last-open field line, and the azimuthal direction.
In addition, $h_1$ and $h_2$ represent the thickness of the main accretion
region and the total gap thickness, respectively.

Instead of solving the real three-dimensional Poisson equation, we divide the gap into many equal divisions in the azimuthal direction and regard each slice of the gap as a two-dimensional two-layer gap. For each division, the solution of the potential in our two-dimensional study is applied. The shape of the $\gamma$-ray spectrum of the division is determined by three parameters: $f$, fractional gap thickness,
$h_1/h_2$, ratio of the thicknesses of the primary and whole region,
and $\rho_1$, the number density in the main acceleration region. 

We define the gap fraction $f$
 measured  on the stellar surface as \cite{2D},
\begin{equation}
f\equiv\frac{h_2(R_s)}{r_p},
\label{def_fm}
\end{equation}
where $R_s$ is the stellar radius, and $r_p(\phi_p)$ is the
polar cap radius. The accelerating electric field $E_{||}$ is proportional to $f^2$. 

The accelerating electric is caused by the deviation of the charge density from the Goldreich-Julian charge density, so we introduce a parameter $g$ to represent this deviation, the $g$ and $\rho$ satisfy the relation, $\rho(x,z,\phi_p)-\rho_{GJ}(x,\phi_p)\sim g(z,\phi_p)\rho_{GJ}(x,\phi_p)$. For each slice of the gap, the deviations of the charge density in the two region, $-g_1$ and $g_2$, and $h_1/h_2$ satisfy
\begin{equation}
(\frac{h_2}{h_1})^2=1+\frac{g_1}{g_2}.
\label{eqn9}
\end{equation}

\section{SIMULATION OF THE ENERGY DEPENDENT LIGHT CURVES}
The first step is to determine the viewing angle and the inclination angle by making the `geometry determined' light curve, whose peaks are due to the caustic effect and almost do not dependent on the energy. 
Our two-dimensional study shows that the fractional size of the gap is small (around 0.16), therefore, in this study, we trace the magnetic field lines of $a=1 \to 0.93$. 
The factor $a$ is used to represent the magnetic field lines at a given layer, where $a=1$ and $a=0$ represent the last-open field lines and magnetic axis, respectively.

The emission direction is calculated in observer's frame \cite{Ta07}. The curvature photon is assumed to be emitted in the direction of the particle motion, which can be described as an combination of the motion along the magnetic field line and the drift motion,
\begin{equation}
\vec{v}=v_p\vec{B}/B + \vec{r}\times\vec{\Omega},
\end{equation}
where $v_p$ is calculated from the condition that $|\vec{v}|=c$. The polar angle to the rotation axis $\zeta$ of the emission direction
 and the pulse phase  $\psi$  are calculated from \cite{Ya97}
\begin{equation}
\left\{
   \begin{array}{ccc}
             \cos{\zeta}=v_z/v\\
             \psi=-\cos^{-1}(v_x/v_{xy})-\vec{r}\cdot\vec{\hat{v}}/R_L
   \end{array}\right.
\end{equation}
If a point in a field line satisfies $|\zeta-\beta|<\varphi(r)$, the radiation of that point can be seen by the observer, where
 $\beta$ is the viewing angle and $\varphi(\vec{r})$ is the solid angle of the radiation.
 
\begin{figure}[t]
\centering
\includegraphics[width=65mm]{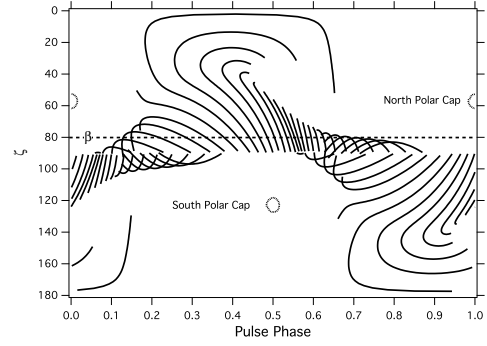}
\caption{The skymap of the radiations outside the null charge surface with inclination angle $\alpha=57^{\circ}$. The $x$-axis is the pulse phase and the $y$-axis is the direction of the radiation. {{}The dashed line is the viewing angle, which is chosen as $\beta=80^{\circ}$}.} \label{skymap}
\end{figure}

We find that the set of
($\alpha,~\beta,~a_{min})=(57^{\circ},~80^{\circ},~0.935$) reproduces
right peak separation of the two peaks. Figure~\ref{skymap} shows
the skymap of the emitted photons from the magnetic surface of $a=0.95$ with
$\alpha=57^{\circ}$ and $\beta=80^{\circ}$.

Then we calculate the $\gamma$-ray spectrum to find proper values of the three parameters, $1-g_1$, $h_1/h_2$ and $f$. We find that the values of these three parameters used in the two-dimensional study can provide a good spectrum. Here we assume the distance to the pulsar is 325~pc.

At last, we calculate the phase resolved spectra under the parameters found above, and integrate the phase resolved spectra to obtain the energy dependent light curves.
The number of the photons
 measured at pulse phases between $\psi_1$ and $\psi_2$ is calculated from
\begin{equation}
N_{\gamma}(E_1, E_2, \psi_1, \psi_2)\propto\int^{E_2}_{E_1}{F_{tot}(E, \psi_1 \leq \psi \leq \psi_2)}dE,
\label{int_spec}
\end{equation}
where the $F_{tot}$ is the phase resolved spectrum.

The calculated pulse profile under constant values of the three parameters in azimuthal direction can not explain the existence of the observed third peak.
The fact that the simple caustic model can not explain the
phase shift of the observed third peak,
forces us to consider more complex structure of the emission region.

\begin{table*}[t]
\begin{center}
\caption{The effects of the distributions of the three parameters}
\begin{tabular}{|c|c|c|c|c|c|}
\hline \textbf{$f(\phi_p)$} & \textbf{$h_1/h_2(\phi_p)$} & \textbf{$\rho_1(\phi_p)$} &
\textbf{The Bump} & \textbf{The Third Peak} & \textbf{The Third Peak}\\
 & & \textbf{($1-g_1$)} & \textbf{$0.1GeV<E<1GeV$} & $1GeV<E<8GeV$ & $8GeV<E<20GeV$
\\
\hline 0.2 & 0.927 & 0.05 & no & no & no \\

\hline 0.2 & Equation~{\ref{h1/h2}} & 0.05 & phase $\sim 0.2-0.3$ & no & no\\

\hline 0.2 & 0.927 & Equation~{\ref{eqn9}} \&~{\ref{rhoave}} & no & phase $\sim 0.26$ & phase $\sim 0.26$ \\

\hline Equation~{\ref{def_f}} & 0.927 & 0.05 & no & no & phase $\sim 0.4$\\

\hline Equation~{\ref{def_f}} & Equation~{\ref{h1/h2}} & Equation~{\ref{eqn9}} \&~{\ref{rhoave}} & phase $\sim 0.2-0.3$ & phase $\sim 0.26$ & phase $\sim 0.33$\\
\hline
\end{tabular}
\label{3dis}
\end{center}
\end{table*}

By the  definition of $f$ in equation~(\ref{def_fm}),
we may choose the form of the azimuthal distribution of $f$ as,
\begin{equation}
f(\phi_p)=\frac{C}{r_p(\phi_p)},
\label{def_f}
\end{equation}
where the $C=0.18r_{p}^{max}$, $r_{p}^{max}$ is the maximum value of the
 polar cap radius, and the factor 0.18 is chosen
by fitting the phase averaged-spectrum.

It is expected that
as the null charge surface is closer
 to the stellar surface, the  number density of the X-ray photons increases
in the gap and the screening region becomes thinner due to higher pair creation rate.
Therefore, we assume the formula of the ratio as
\begin{equation}
\frac{h_1}{h_2}(\phi_p)=B_1+B_2\frac{1/r_{null}(\phi_p)-1/r_{null}^{max}}{1/r_{null}^{min}-1/r_{null}^{max}},
\label{h1/h2}
\end{equation}
By fitting
the spectral shape, we obtain $B_1=0.89$ and $B_2=0.09$.

With the two-dimensional two-layer model,
we found that the phase-averaged spectra for most of the $\gamma$-ray pulsars
can be reproduced  by the averaged charge density of
$\bar{\rho}_0\equiv [h_1\rho_1+(h_2-h_1)\rho_2]/h_2\sim 0.5$.
In three-dimensional magnetosphere, the actual $E_{\perp}(\phi_p)$ can make the particles in different $\phi_p$-cell 
to drift into other $\phi_p$-cell via the effect of $E_{\perp}\times B$.
As a result, the actual average density is
\begin{equation}
\bar{\rho}(\phi_p)=\bar{\rho}_0\frac{f(\phi_p+\Delta{\phi_p})}{f(\phi_p)},
\label{rhoave}
\end{equation}
where the $\bar{\rho}(\phi_p)=N(\phi_p)/f(\phi_p)$ is used and $\bar{\rho}_0=0.5$ is the averaged charge density without
drift motion. Using  the relationship between $g_1$, $g_2$ and $h_1/h_2$ given by equation~(\ref{eqn9}),
the distribution of $1-g_1$ can be obtained. 

We expected that the particles' displacement due to
the drift motion becomes more important on the magnetic field line that has
a smaller radial distance to the  null charge surface,
 because the particles run
longer distance in the outer gap. To take into this effect, we assume the formula of displacement as
\begin{equation}
\Delta{\phi_p}=F\frac{1/r_{null}(\phi_p)-1/r_{null}^{max}}{1/r_{null}^{min}-1/r_{null}^{max}},
\end{equation}
where $F$ is a fitting parameter, which is chosen as -28$^{\circ}$.

\begin{figure}[t]
\centering
\includegraphics[width=86mm]{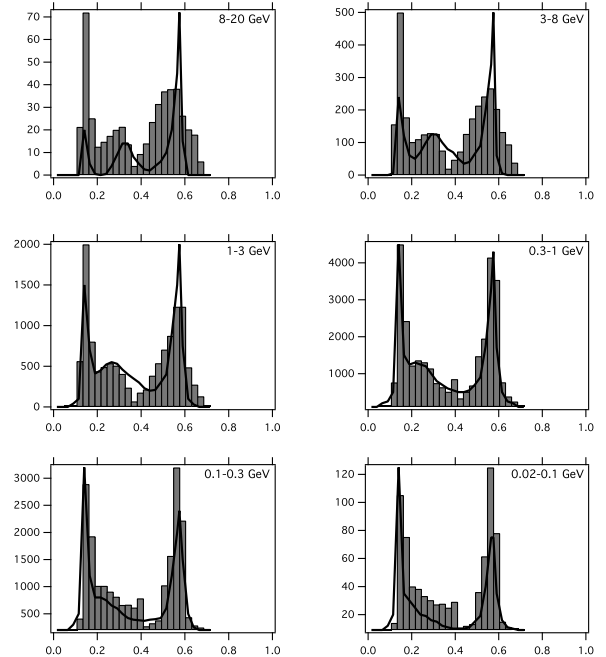}
\caption{The energy dependent light curves (histogram) with the distributions of $1-g_1$, $h_1/h_2$ and $f$, provided by the three-dimensional outer gap. The solid lines are the observed light curves from Fermi-LAT \cite{obs_Vela}.} \label{edlc_3dis}
\end{figure}

\begin{figure}[t]
\centering
\includegraphics[width=65mm]{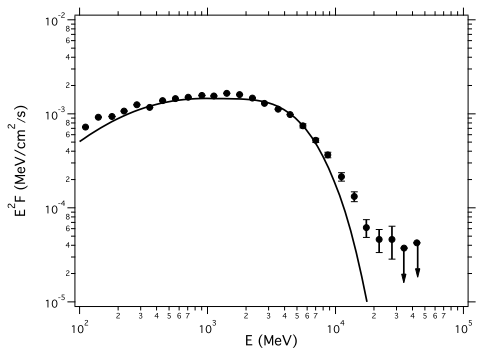}
\caption{The phase averaged spectrum with the distributions of $1-g_1$, $h_1/h_2$ and $f$, comparing with the observed data (circle) from Fermi-LAT \cite{obs_Vela}} \label{spec}
\end{figure}

\begin{figure}
\centering
\includegraphics[width=80mm]{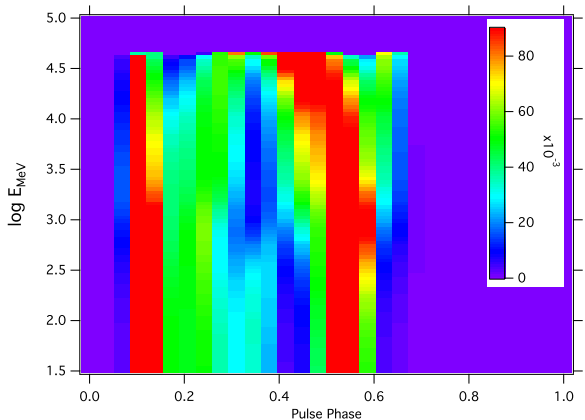}
\caption{Intensity map in the pulse phase and energy plane. The color represents the percentage of the number of the photons of certain interval of pulse phase in the total number of photons of certain interval of energy.} 
\label{colmap}
\end{figure}

Table~{\ref{3dis}} summaries the effects of the distributions of the parameters on the energy dependent light curves.
Figure~\ref{edlc_3dis} shows the pulse profiles calculated
by taking into account the azimuthal distributions $f$, $h_1/h_2$ and $1-g_1$. The corresponding phase averaged spectrum is shown in Figure~{\ref{spec}}. Figure~\ref{colmap} is the intensity map in the pulse phase and the energy plane. The color
represents the scaled number of the photons at the certain interval of the pulse phase.
We find a moving third peak from the results.

\begin{figure}
\begin{minipage}[c]{0.5\linewidth}
\centering
\includegraphics[width=35mm]{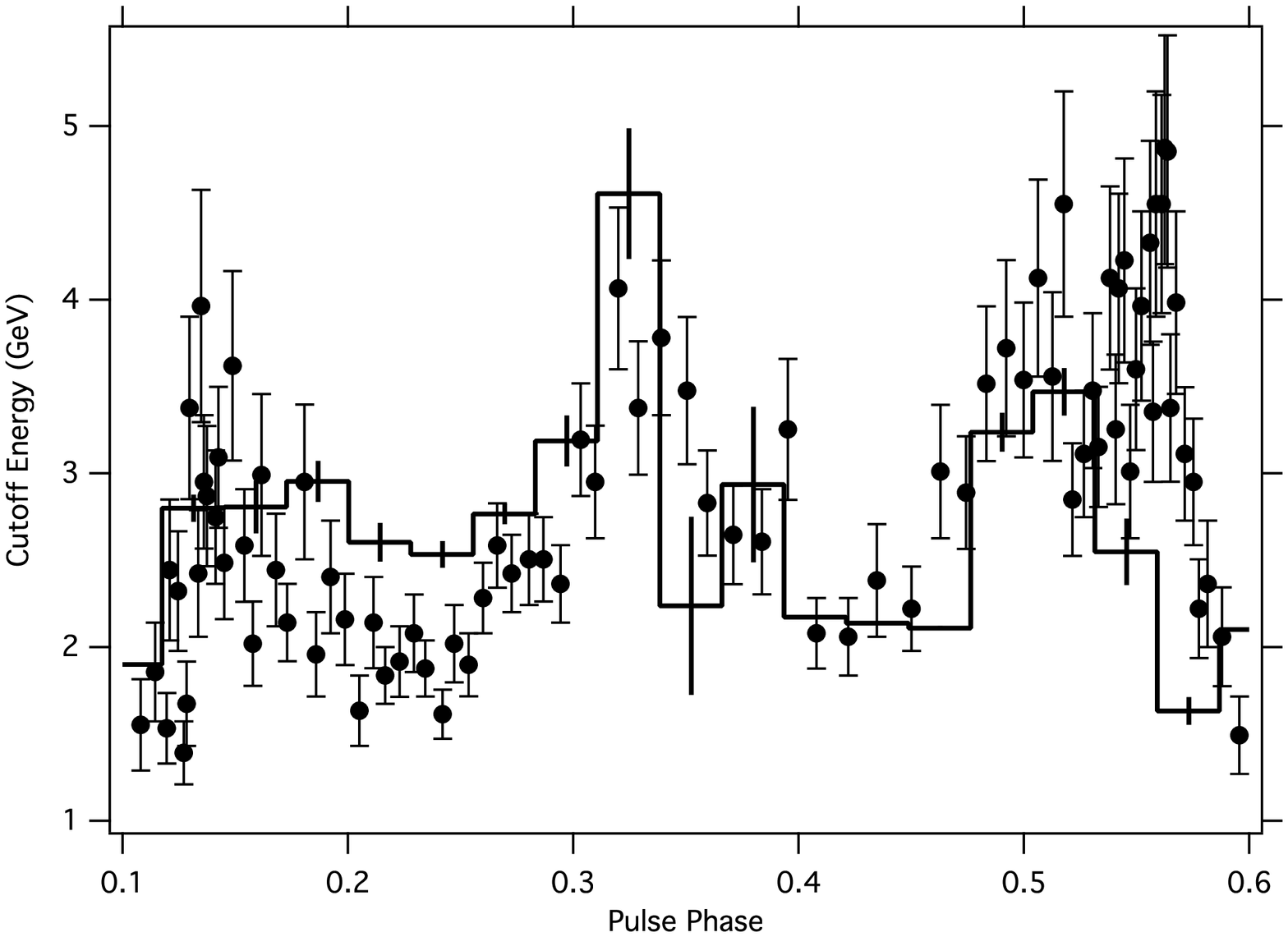}
\end{minipage}%
\begin{minipage}[c]{0.5\linewidth}
\centering
\includegraphics[width=35mm]{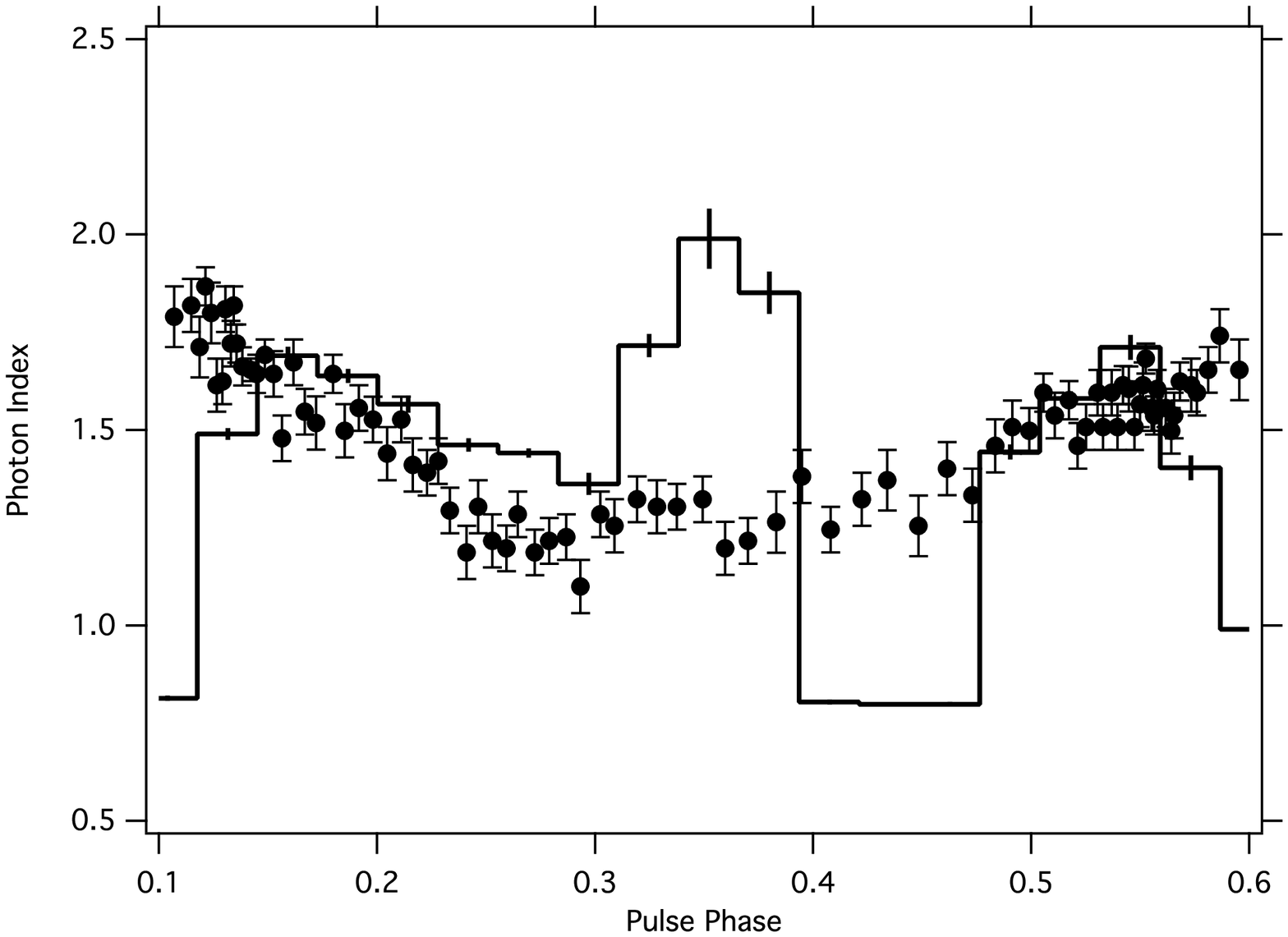}
\end{minipage}
\caption{The cutoff energies and photon indices of the phase-resolved spectra of different pulse phases, comparing with the observed data (circle) \cite{obs_Vela}.}
\label{EcutGamma}
\end{figure}

We also fit the calculated phase resolved spectra with power law plus exponential cut-off form.
 Figure~\ref{EcutGamma} shows
the cut-off energy and the photon index, as functions of the
pulse phase.


\section{THE REASON FOR THE THIRD PEAK}
The reason for the third peak and its shift is as follows.
In the light curves, when energy lower than 1~GeV, the distributions
of the thickness ratio $h_1/h_2$ makes a third-peak structure at $\sim0.2$
pulse phase, because $h_1/h_2$
affects the emissivity in the screening region,
which mainly produces the curvature photons of the energy less than
 1~GeV. In the energy bands higher than 1~GeV, on the other hand,
 the azimuthal distributions of the fractional thickness, $f$, and the number
density, $1-g_1$, are more important, and they produce the third  peak
at $\sim 0.3-0.35$ pulse phase by providing higher electric field and more particles, respectively.
Consequently, the differences in the standing phases of the
third peak due to the distributions of $h_1/h_2$, $f$ and $\rho_1$(or$1-g_1$)
 produce the shift of the third peak with the photon energy.
 
Why is the Vela Pulsar so special? This is because: 1) It has a thin gap to make sure that its observable radiation contains both the contributions of the two layers. 2) Its $\gamma$-ray radiation is dominated by curvature photons that keep the effects of azimuthal structure of the gap visible. 3) Two-dimensional study shows that the shape of the spectrum is very sensitive to $\rho_1(\phi_p)$ and $\frac{h_1}{h_2}(\phi_p)$, when $\rho_1$ and $h_1/h_2$ get close to 0 and 1, respectively. This makes the effects of azimuthal structure of the gap obvious. 4) The inclination angle also determines the azimuthal structure of the gap.

\section{CONCLUSION}
The third peak of Vela Pulsar can not be explained by the caustic effect determined by the magnetic field structure. This peak is due to the azimuthal distributions of the
fractional gap thickness ($f$),
the ratio of the thicknesses of the primary and whole region ($h_1/h_2$),
and the number density in the main acceleration region ($\rho_1$). The distributions of $\rho_1$ and $f$ make third-peak-like structure
in the bridge region of light curve above 1 GeV, while the distribution of
$h_1/h_2$ makes a bump in the bridge region of the light curves below 1 GeV.
The phases of the third peaks caused by the azimuthal distributions of
$h_1/h_2$, $\rho_1$ and $f$ are different from each other. Consequently,
the differences in the phases produce the shift of the combined third peak with the photon energy.

\bigskip 
\begin{acknowledgments}
We thank L. Zhang for useful discussion. This work is supported by a GRF grant of the the Hong Kong SAR Government entitled ``Gamma-ray Pulsars'' (HKU700911P).
\end{acknowledgments}

\bigskip 

\end{document}